\documentclass[conference]{IEEEtran}
\usepackage[shortlabels]{enumitem}
\usepackage[nospace,noadjust]{cite}
\usepackage{float}
\usepackage{graphicx} 
\usepackage{epsfig} 
\usepackage{amsmath,amsthm,amssymb}  
\usepackage{float}
\usepackage{cases,setspace,adjustbox}
\usepackage{array}
\usepackage[update,prepend]{epstopdf}
\usepackage{eqparbox}
\usepackage{url}
\usepackage{caption, subcaption}
\usepackage{xcolor}
\usepackage{mathtools}
\usepackage{bbm}
\usepackage{multirow}
\usepackage{multicol}
\usepackage{algorithm,algorithmic}

\usepackage[acronym]{glossaries}
\usepackage{comment}
\usepackage{xcolor}
\newcommand*\circled[1]{\kern-2.5em%
  \put(0,4){\color{white}\circle*{11}}\put(0,4){\circle{11}}%
  \put(-3,0){\color{black}\large#1}~~}
\setacronymstyle{long-short}
\newacronym{3gpp}{3GPP}{3rd Generation Partnership Project}
\newacronym{5G}{5G}{Fifth-generation}
\newacronym{6G}{6G}{sixth-generation}
\newacronym{c2}{C2}{command-and-control}

\newacronym{ack}{ACK}{Acknowledgement}
\newacronym{AR}{AR}{Augmented Reality}
\newacronym{ai}{AI}{Artificial Intelligence}

\newacronym{bler}{BLER}{Block Error Rate}

\newacronym{ci}{CI}{close-in}
\newacronym{cdf}{CDF}{cumulative distribution function}
\newacronym{ccdf}{CCDF}{complementary cumulative distribution function}
\newacronym{cdl}{CDL}{clustered delay line}
\newacronym{comp}{CoMP}{Coordinated Multi-Point}
\newacronym{cqi}{CQI}{Channel Quality Indicator}
\newacronym{csi}{CSI}{channel state information}

\newacronym{d2d}{D2D}{Device-to-Device}
\newacronym{dc}{DC}{dual connectivity}
\newacronym{dcs}{DCS}{dynamic cell sizing}
\newacronym{dci}{DCI}{Downlink Control Information}
\newacronym{dft}{DFT}{Discrete Fourier Transform}
\newacronym{dl}{DL}{downlink}
\newacronym{dsrc}{DSRC}{Dedicated short-range communications}
\newacronym{dss}{DSS}{Dynamic Spectrum Sharing}
\newacronym{DU}{DU}{Distributed Unit}

\newacronym{edf}{EDF}{Earliest Deadline First}
\newacronym{ekf}{EKF}{extended Kalman filtering}
\newacronym{ets}{ETS}{Exponential Smoothing}
\newacronym{embb}{eMBB}{enhanced mobile broadband}
\newacronym{enb}{eNB}{Evolved Node-B}
\newacronym{epc}{EPC}{Evolved Packet Core}
\newacronym{emtc}{eMTC}{enhanced MTC}
\newacronym{ewma}{EWMA}{exponentially weighted moving average}

\newacronym{fcfs}{FCFS}{First-Come-First-Served}
\newacronym{fdd}{FDD}{Frequency Domain Duplex}
\newacronym{fdm}{FDM}{Frequency Domain Multiplexing}
\newacronym{fifo}{FIFO}{first-in-first-out}
\newacronym{firstnet}{FirstNet}{First Responder Network Authority}
\newacronym{fspl}{FSPL}{free-space path loss}
\newacronym{fpga}{FPGA}{Field-Programmable Gate Array}

\newacronym{gan}{GAN}{Generative Adversarial Network}
\newacronym{gps}{GPS}{Global Positioning System}
\newacronym{gnb}{gNB}{Next Generation Node-B}
\newacronym{gbps}{Gb/s}{gigabits per second}
\newacronym{gbr}{GBR}{guaranteed bitrate}

\newacronym{harq}{HARQ}{Hybrid Automatic Repeat reQuest}
\newacronym{haps}{HAPS}{High-Altitude Platform Stations}
\newacronym{hq}{HQ}{headquarters}
\newacronym{hol}{HoL}{head-of-line}

\newacronym{ici}{ICI}{inter-cell interference}
\newacronym{icmp}{ICMP}{Internet Control Message Protocol}
\newacronym{iot}{IoT}{Internet of Things}
\newacronym{iiot}{IIoT}{Industrial Internet of Things}
\newacronym{iomt}{IoMT}{Internet of Medical Things}
\newacronym{iqr}{IQR}{interquartile range}
\newacronym{irs}{IRS}{Intelligent Reflecting Surfaces}
\newacronym{its}{ITS}{Intelligent Transport Systems}
\newacronym{itu}{ITU}{International Telecom Union}

\newacronym{kpi}{KPI}{Key Performance Indicator}
\newacronym{kpis}{KPIs}{Key Performance Indicators}

\newacronym{los}{LoS}{line-of-sight}
\newacronym{lte}{LTE}{Long Term Evolution}
\newacronym{lteapro}{LTE-A~Pro}{LTE Advanced Pro}
\newacronym{lstm}{LSTM}{Long Short-Term Memory}

\newacronym{mac}{MAC}{Medium Access Control}
\newacronym{mcs}{MCS}{Modulation and Coding Scheme}
\newacronym{mcptt}{MCPTT}{Mission-Critical Push-to-Talk}
\newacronym{mimo}{MIMO}{Multiple-Input Multiple-Output}
\newacronym{mmimo}{mMIMO}{massive Multiple-Input Multiple-Output}
\newacronym{mmtc}{mMTC}{massive machine-type communication}
\newacronym{mtc}{MTC}{Machine-Type Communication}
\newacronym{mlp}{MLP}{Multilayer Perceptron}
\newacronym{ma}{MA}{Moving Average}
\newacronym{mm}{MM}{Moving Median}
\newacronym{ml}{ML}{Machine Learning}
\newacronym{mmwave}{mmWave}{millimeter-wave}

\newacronym{nbiot}{NB-IoT}{Narrowband-Internet of Things}
\newacronym{nr}{NR}{New Radio}
\newacronym{ntia}{NTIA}{National Telecommunications and Information Administration}
\newacronym{ntn}{NTN}{Non-Terrestrial Networks}
\newacronym{nlos}{NLoS}{non-line-of-sight}

\newacronym{oran}{O-RAN}{Open Radio Access Network}
\newacronym{owl}{OWL}{Online Watcher of LTE}

\newacronym{papf}{PA-PF}{Priority-Aware Proportional Fair}
\newacronym{prb}{PRBs}{Physical Resource Blocks}
\newacronym{pf}{PF}{Proportional Fair}
\newacronym{psc}{PSC}{Public Safety Communications}
\newacronym{pscch}{PSCCH}{Physical Sidelink Control Channel}
\newacronym{psfch}{PSFCH}{Physical Sidelink Feedback Channel}
\newacronym{pssch}{PSSCH}{Physical Sidelink Shared Channel}

\newacronym{qos}{QoS}{Quality of Service}

\newacronym{ran}{RAN}{Radio Access Network}
\newacronym{rApp}{rApp}{RAN Application}
\newacronym{rans}{RANs}{Radio Access Networks}
\newacronym{rat}{RAT}{Radio Access Technology}
\newacronym{re}{RE}{Resource Element}
\newacronym{rf}{RF}{radio-frequency}
\newacronym{ric}{RIC}{RAN Intelligent Controller}
\newacronym{relu}{ReLU}{Rectified Linear Unit}
\newacronym{rl}{RL}{Reinforcement Learning}
\newacronym{rmse}{RMSE}{root-mean-square error}
\newacronym{rnn}{RNN}{Recurrent Neural Network}
\newacronym{rr}{RR}{Round Robin}
\newacronym{rssi}{RSSI}{Received Signal Strength Indicator}
\newacronym{rsrp}{RSRP}{Reference Signal Received Power}
\newacronym{RU}{RU}{Radio Unit}
\newacronym{rt}{RT}{real-time}
\newacronym{rtt}{RTT}{round-trip time}

\newacronym{scm}{SCM}{Spatial Channel Model}
\newacronym{sdr}{SDR}{Software Defined Radio}
\newacronym{sem}{SEM}{Statistical Ensemble Method}
\newacronym{sl}{SL}{sidelink}
\newacronym{smo}{SMO}{Service and Management Orchestration}
\newacronym{snr}{SNR}{Signal-to-Noise Ratio}
\newacronym{sinr}{SINR}{Signal-to-Interference-plus-Noise Ratio}
\newacronym{ssb}{SSB}{Synchronization Signal Block}

\newacronym{td3}{TD3}{Twin Delayed Deep Deterministic Policy Gradient}
\newacronym{tdd}{TDD}{Time Domain Duplex}
\newacronym{tdm}{TDM}{Time Domain Multiplexing}
\newacronym{timegan}{TimeGAN}{Time-Series Generative Adversarial Network}
\newacronym{tr}{TR}{Technical Report}
\newacronym{thz}{THz}{terahertz}
\newacronym{TIP}{TIP}{Telecom Infra Project}
\newacronym{tti}{TTI}{Transmission Time Interval}

\newacronym{urllc}{URLLC}{ultra-reliable low-latency communication}
\newacronym{uav}{UAV}{Unmanned Aerial Vehicle}
\newacronym{ue}{UE}{User Equipment}
\newacronym{ul}{UL}{uplink}
\newacronym{upa}{UPA}{uniform planar array}

\newacronym{v2v}{V2V}{Vehicle-to-Vehicle}
\newacronym{v2x}{V2X}{Vehicle-to-Everything}
\newacronym{VR}{VR}{Vitual Reality}
\newacronym{vnf}{VNFs}{Virtualized Network Functions}
\newacronym{vm}{VMs}{Virtual Machines}

\newacronym{wlan}{WLAN}{Wireless Local Area Network}
\newacronym{wssus}{WSSUS}{wide-sense stationary uncorrelated scattering}
\newacronym{xApp}{xApp}{eXtended application}
\newacronym{xr}{XR}{Extended Reality}
\begin{document}
\title{Decomposing URLLC Deadline Failures in Beam-Managed 6G Networks: Implications for Mission-Critical Wireless Systems}
\author{Webert Montlouis, Fellow, IEEE~\IEEEmembership{Member,~IEEE,}
        and Sneihil Gopal, Senior Member, IEEE\\
Emails: \{webert\_montlouis, sneihil.gopal\}@ieee.org}
\maketitle
\begin{abstract}
Mission-critical wireless systems require ultra-reliable low-latency communications (URLLC), which sixth-generation (6G) millimeter-wave (mmWave) beam-managed networks are designed to provide. However, the reported URLLC deadline miss rate often masks the specific causes of packet failures. This paper introduces a framework that categorizes missed URLLC packets into three causes: queueing-limited (Q), transmission-limited (T), and beam-tracking-limited (B). Our analysis across ten scenarios with antenna counts from $\mathbf{64}$ to $\mathbf{4096}$ reveals three key findings. First, the dominant cause of missed packets shifts from Q to B when the antenna count exceeds roughly $\mathbf{1825}$. Therefore, operators of hyper-massive arrays should focus on improving beam-tracking accuracy to reduce the miss rate. Second, the beam-sweep period ($T_{\mathrm{sw}}$) significantly affects B-misses, while Q and T remain relatively unchanged. Reducing $T_{\mathrm{sw}}$ from $\mathbf{20}$\,ms to $\mathbf{10}$\,ms can cut the B-misses by more than half, without requiring a redesign of the scheduler. Third, Q-misses reach a cumulative distribution function (CDF) of approximately $\mathbf{0.98}$ by $\mathbf{0.8}\,$ms, so modest latency-budget relaxation can accommodate most overruns. 
In contrast, B-misses remain unsaturated at $\mathbf{12}\,$ms, making them difficult to accommodate within practical application-layer latency limits. Overall, this framework transforms an ambiguous aggregate miss-rate metric into a clear three-way diagnostic tool that directly informs remedial actions for network operators. 
\end{abstract}

\section{Introduction}
\label{sec:intro}
\let\thefootnote\relax\footnotetext{This work used Stampede3 at TACC through allocation  ASC160068  from the Advanced Cyberinfrastructure Coordination Ecosystem: Services \& Support (ACCESS) program, which is supported by U.S. National Science Foundation grants $\#2138259, \#2138286, \#2138307, \#2137603$, and $\#2138296$.}Mission-critical wireless systems require deterministic latency guarantees for \gls{c2}, sensing, and autonomous coordination. Although \gls{6G} \gls{mmwave} beam-managed networks provide the capacity needed for these applications at scale~\cite{popovski2018,3gpp38913}, \gls{urllc} deadline misses are typically reported as a single aggregate metric. This approach obscures the underlying failure mechanisms. A miss caused by scheduling congestion requires a different remedy than one caused by beam misalignment on a moving platform, which is distinct from a miss due to poor channel quality. Without identifying the causes of these misses, aggregate miss-rate measurements do not provide guidance for targeted remediation~\cite{survey_sched_urllc}. This issue is particularly acute in military contexts. For instance, a ground commander noticing increasing deadline-miss rates from a \gls{uav} command link cannot determine, based on the aggregate metric alone, whether to request additional spectrum, reposition the ground station, update the beam codebook, or tighten the beam-sweep schedule. Each decision is linked to a different underlying cause, and committing resources to the wrong solution can waste both time and capacity in environments where both are scarce and potentially contested.

Previous work on \gls{mmwave} beam management has focused on codebook design, beam tracking, and sweep-overhead tradeoffs relevant to this setting~\cite{giordani2019,xue2023}. Research on \gls{urllc}~\cite{anand2020,survey_sched_urllc} scheduling and \gls{mmimo} scaling~\cite{larsson2014,bjornson2025gigantic} provides complementary foundations, but none of this work breaks down the \gls{urllc} miss rate by proximate cause. As a result, operators are unable to identify which system layer to address when miss rates increase. Furthermore, even an optimal scheduler cannot eliminate beam-tracking failures, so the dominant cause cannot be inferred solely from scheduler performance. To illustrate this point, the decomposition is evaluated alongside six schedulers that span the policy range from \gls{fcfs} to \gls{ml}-augmented \gls{edf}. This evaluation demonstrates that cause-specific remediation cannot be replaced by simply selecting a different scheduler, regardless of the policy used. 

This paper addresses the decomposition gap through four main contributions. First, it develops a decomposition framework with formal definitions that partitions \gls{urllc} deadline misses into three proximate-cause categories: queueing-limited (Q), transmission-limited (T), and beam-tracking-limited (B). The framework also includes a low-overhead per-packet classifier that can be incorporated into deadline-aware \gls{mmwave} simulators with negligible instrumentation cost. Second, an empirical evaluation across ten deployment scenarios, ranging from small tactical nodes to hyper-massive \gls{uav} coordination platforms, demonstrates that the dominant failure mechanism shifts from Q to B as the antenna count exceeds approximately $1825$. This crossover has direct design implications for military \gls{mmwave} systems, indicating that arrays beyond this point benefit increasingly from improved beam tracking rather than further scheduler optimization. Third, a sensitivity analysis identifies the sweep period, $T_\mathrm{sw}$, as a primary factor governing the B-component, while the Q and T components remain largely unchanged as $T_\mathrm{sw}$ varies. Reducing $T_\mathrm{sw}$ from $20\,$ms to $10\,$ms decreases B-misses by more than half without requiring scheduler redesign, supporting a low-cost, mobility-adaptive sweep policy for \gls{uav} and vehicular platforms. 
Finally, the per-cause overrun-delay analysis shows that Q-misses are concentrated within sub-millisecond overruns, whereas the B-miss distribution remains unsaturated at the $12\,$ms limit of the evaluated range, making application-layer accommodation increasingly impractical. These results identify beam-tracking failures as an important limiting mechanism for safety-critical \gls{c2} and collision-avoidance links under the studied conditions. 

The remainder of the paper is organized as follows. Section~\ref{sec:related} reviews related work. Section~\ref{sec:system} summarizes the system model. Section~\ref{sec:framework} presents the decomposition framework and classifier. Section~\ref{sec:results} describes the simulation setup and presents empirical results. Section~\ref{sec:implications} maps findings to cause-specific remedial actions. Section~\ref{sec:limitations} addresses limitations and Section~\ref{sec:conclusion} concludes.
\section{Related Work}
\label{sec:related}
Scheduling and resource allocation for \gls{urllc} have received significant attention in the \gls{5G} and pre-\gls{6G} literature. Authors in~\cite{anand2020} investigate the joint scheduling of \gls{urllc} and \gls{embb} traffic, characterizing the trade-offs between throughput and reliability. A comprehensive survey of scheduling algorithms for \gls{5G} \gls{urllc} and their applicability to \gls{6G} is provided in~\cite{survey_sched_urllc}. The classic \gls{edf} result from~\cite{liu1973} continues to serve as a foundation for deadline-aware scheduling in hard real-time systems. Beam management for \gls{mmwave} systems is thoroughly reviewed in~\cite{giordani2019} and~\cite{xue2023}, addressing codebook design, beam tracking, and the trade-offs between sweep overhead and beam coherence time. The fundamentals of signal processing and beamforming gains for \gls{mmwave} MIMO and \gls{mmimo} systems are characterized in~\cite{heath2016} and~\cite{larsson2014}, highlighting a trend towards narrower beamwidth that underlies the observed growth in performance metrics. Authors in~\cite{bjornson2025gigantic} extend this analysis to gigantic and hyper-\gls{mmimo} configurations targeting \gls{6G} deployments, establishing that arrays with 
a large number of antennas operate in a qualitatively different beamwidth regime, where angular tracking accuracy becomes the primary constraint on link reliability.

For mobile platforms, beam tracking using the \gls{ekf}~\cite{ekf_beam_tracking} has been shown to reduce the probability of misalignment in vehicular and \gls{uav} scenarios by leveraging angular-state prediction. The authors in~\cite{evangelista2022} analyze user-plane latency and reliability for \gls{mmwave} \gls{mmimo} under grant-free \gls{urllc} access, providing complementary per-class reliability results, although they do not decompose missed connections by cause. To the best of our knowledge, this paper is the first to analyze the \gls{urllc} miss rate based on specific proximate causes across a systematic multi-scenario evaluation, identifying that approximately $1825$ antennas represent a threshold at which beam-tracking failures exceed scheduling congestion as the dominant reason for misses. This finding allows for the cause-specific remediation strategies outlined in Section~\ref{sec:implications}.

\section{System Model}
\label{sec:system}
The framework outlined in~\cite{giordani2019} is employed by the system. A base station equipped with a uniform planar array of $M$ antennas manages $B$ analog beams while serving $U$ users located within a cell of radius $R$. The channel utilizes the \gls{3gpp} TR~$38.901$ urban-microcell \gls{los} path-loss model at a carrier frequency of $f_c = 28$\,GHz~\cite{3gpp38901}. Beam sweeping is executed with a sweep period of $T_{\mathrm{sw}} = 20$\,ms (the default setting), and beam tracking occurs at a resolution of $T_{\mathrm{tr}} = 1$\,ms. 

Traffic is categorized into three distinct classes: \gls{urllc}, \gls{embb}, and \gls{mmtc}, with hard deadlines of $1$, $4$, and $10\,\mathrm{ms}$, respectively. In the context of mission-critical applications addressed by this paper, \gls{urllc} traffic includes \gls{c2} messages, sensor telemetry, and time-sensitive targeting information. A \gls{urllc} packet $p$, created at time $t^{\mathrm{arr}}$, is assigned an absolute deadline of $t^{\mathrm{dl}} = t^{\mathrm{arr}} + D_{\mathrm{URLLC}}$, where $D_{\mathrm{URLLC}} = 1\,\mathrm{ms}$. The packet is considered missed if $t^{\mathrm{done}} > t^{\mathrm{dl}}$ or if $t^{\mathrm{done}} = \infty$ for packets that remain undelivered. The beam-managed channel takes into account path loss as specified in TR~$38.901$, along with log-normal shadowing, Rayleigh fading, and a $12$\,dB misalignment penalty that is applied when the tracked beam diverges from the user's actual angular position. The probability of misalignment is influenced by user velocity, cell radius, $T_{\mathrm{sw}}$, and $B$, following~\cite{giordani2019}. This phenomenon illustrates the fundamental trade-off between array gain (which increases with $M$) and the beam's susceptibility to angular drift (which also increases with $M$ as beamwidths reduce proportionally to $1/\sqrt{M}$).

Six schedulers are compared to show that the overall miss rate is an inadequate measure, regardless of the scheduling policy used. \textbf{\gls{edf}} selects packets based on the nearest absolute deadline, following the classical hard-real-time policy in~\cite{liu1973}. In contrast, \textbf{\gls{pf}} allocates resources to balance throughput without considering deadlines~\cite{kushner2004}. \textbf{\gls{rr}} serves users in a fixed rotation, treating all traffic equally, while \textbf{\gls{fcfs}} processes packets in their arrival order, offering a worst-case baseline. 
\textbf{Hybrid \gls{ml}-\gls{edf}} enhances \gls{edf} urgency scores using channel state and queue depth while retaining deadline awareness. Lastly, \textbf{Weighted \gls{ml}-\gls{edf}} uses a learned weight vector to optimize both deadline adherence and resource utilization, making it the strongest scheduler evaluated.
\section{Decomposition Framework}
\label{sec:framework}
To determine the proximate cause of each missed \gls{urllc} packet, we establish three distinct and comprehensive failure classes along with a low-overhead per-packet classifier that assigns each miss to one specific class immediately upon detection. This framework requires only three integer counters and access to packet timestamps and beam-alignment state, which most discrete-event beam-management simulators already maintain. Importantly, the three causes are observationally indistinguishable at the aggregate level: a packet that fails to meet its deadline due to beam misalignment results in the same outcome metric as one that misses because of \gls{sinr}-induced retransmissions or scheduling congestion. However, each cause requires a fundamentally different remedial action. Distinguishing them requires per-packet information that the overall miss rate overlooks, specifically the scheduling timestamp, the transmission-completion timestamp, and the beam-alignment status at the time of transmission. This makes the classifier an essential instrumentation addition rather than merely a retrospective analysis of existing data.

\subsection{Three Proximate-Cause Categories}
A \gls{urllc} packet can encounter three failure mechanisms before it is delivered or expires:
\begin{itemize}[leftmargin=*]
    \item \textbf{Queueing:} the packet remains in the per-user queue until selected by the scheduler.
    \item \textbf{Transmission:} the packet is sent on a beam, possibly requiring multiple attempts due to \gls{sinr}-induced block errors.
    \item \textbf{Beam alignment:} the beam must remain angularly aligned with the user throughout the transmission attempt.
\end{itemize}

A missed packet is attributed according to its scheduling, transmission, and beam-alignment state. Using the packet deadline $t^{\mathrm{dl}}$ defined in Section~\ref{sec:system}, let $t^{\mathrm{sched}}$ denote the time at which packet $p$ is selected for transmission and $t^{\mathrm{tx\text{-}end}}$ its transmission completion time. Let $\theta_u(\tau)$ and $\hat{\theta}_{b(u)}(\tau)$ denote the user and tracked-beam angles, and $\theta_{\mathrm{bw}}$ the half-power beamwidth. Define $\mathrm{misaligned}(p)\in\{0,1\}$ as $1$ if $|\theta_u(\tau)-\hat{\theta}_{b(u)}(\tau)|>\theta_{\mathrm{bw}}$ for any $\tau$ during transmission, and $0$ otherwise.

\textbf{Queueing-limited (Q):}
A packet is queueing-limited if it is not scheduled before its deadline:
\begin{equation}
t^{\mathrm{sched}} > t^{\mathrm{dl}}.
\label{eq:q_miss}
\end{equation}
Q-misses indicate that scheduling or resource allocation is the limiting factor and can be mitigated through urgency-aware scheduling, reserved beam slots, or additional spectrum.

\textbf{Transmission-limited (T):}
A packet is transmission-limited if it is scheduled on time but completes after the deadline without beam misalignment:
\begin{equation}
\begin{aligned}
t^{\mathrm{sched}} &\leq t^{\mathrm{dl}}
\;\land\;
t^{\mathrm{tx\text{-}end}} > t^{\mathrm{dl}}
\;\land\;
\mathrm{misaligned}(p)=0.
\end{aligned}
\label{eq:t_miss}
\end{equation}
T-misses reflect poor channel quality or retransmissions and can be mitigated by increasing array gain or bandwidth, or reducing cell radius. Under adversarial jamming, the T-component could increase, motivating link-protection mechanisms.

\textbf{Beam-tracking-limited (B):}
A packet is beam-tracking-limited if it is scheduled on time but completes after the deadline with beam misalignment:
\begin{equation}
\begin{aligned}
t^{\mathrm{sched}} &\leq t^{\mathrm{dl}}
\;\land\;
t^{\mathrm{tx\text{-}end}} > t^{\mathrm{dl}}
\;\land\;
\mathrm{misaligned}(p)=1.
\end{aligned}
\label{eq:b_miss}
\end{equation}
B-misses are particularly relevant to mobile platforms with high antenna counts, where narrower beams increase sensitivity to angular drift.

Let $N_Q$, $N_T$, and $N_B$ denote the numbers of \gls{urllc} misses attributed to Q, T, and B, and let $N_{\mathrm{URLLC}}^{\mathrm{miss}}$ and $N_{\mathrm{URLLC}}^{\mathrm{total}}$ denote the missed and total \gls{urllc} packet counts, respectively. Under the proposed proximate-cause attribution, the three categories are mutually exclusive and exhaustive:
\begin{equation}
N_{\mathrm{URLLC}}^{\mathrm{miss}} = N_Q + N_T + N_B.
\label{eq:decomposition}
\end{equation}
The miss rate is additively decomposed as
$M_{\mathrm{URLLC}} = M_Q + M_T + M_B$, where
$M_x = N_x/N_{\mathrm{URLLC}}^{\mathrm{total}}$ for
$x\in\{Q,T,B\}$. Each component therefore identifies the corresponding engineering lever with minimal additional instrumentation.

\subsection{Classifier Algorithm}
Algorithm~\ref{alg:classifier} implements the preceding definitions using the packet timestamps and misalignment indicator defined above. Most discrete-event beam-management simulators already track these quantities. The overhead is limited to three integer counters and one Boolean check per missed packet, which is negligible compared with packet-event processing.
\begin{algorithm}
\caption{Per-Packet Miss Classifier}
\label{alg:classifier}
\begin{algorithmic}[1]
\REQUIRE Packet $p$; timestamps $t^{\mathrm{dl}}$, 
$t^{\mathrm{sched}}$, $t^{\mathrm{tx\text{-}end}}$; 
indicator $\mathrm{misaligned}(p)$

\IF{$t^{\mathrm{tx\text{-}end}} \leq t^{\mathrm{dl}}$}
    \STATE $\mathrm{cause}(p) \leftarrow \mathrm{hit}$
\ELSIF{$t^{\mathrm{sched}} > t^{\mathrm{dl}}$}
    \STATE $\mathrm{cause}(p) \leftarrow Q;\quad N_Q \leftarrow N_Q + 1$
\ELSIF{$\mathrm{misaligned}(p) = 1$}
    \STATE $\mathrm{cause}(p) \leftarrow B;\quad N_B \leftarrow N_B + 1$
\ELSE
    \STATE $\mathrm{cause}(p) \leftarrow T;\quad N_T \leftarrow N_T + 1$
\ENDIF
\end{algorithmic}
\end{algorithm}
For runtime classification on deployed hardware, the misalignment predicate can be substituted with a proxy for the received signal quality drop. Specifically, a sudden drop in either \gls{rsrp} or \gls{sinr} at the start of a transmission attempt, falling below a threshold calibrated to a $12$\,dB misalignment penalty, can serve as a measurable proxy for beam misalignment. This removes the need for ground-truth angular state and provides a practical proxy for future deployment on real mmWave hardware without additional angular-sensing infrastructure. 
\section{Results}
\label{sec:results}
The decomposition is applied across the ten scenarios outlined in Table~\ref{tab:scenarios}. This set of scenarios spans three operational regimes: small-to-medium tactical nodes (S1--S3, $M \le 256$, $U=100$), representative of forward operating bases and headquarters elements; high-density \gls{uav} swarm hubs (S4--S5, $M=256$, $U=1000$); and extreme-density urban \gls{c2} or hyper-massive array deployments (S6--S10, $M \ge 1024$, $U=1000$). Each scenario is executed for $100$ Monte Carlo trials, using independent random seeds. 
The cell-sizing mode employs \gls{dcs} with expansion and contraction load thresholds \(L_{\mathrm{exp}}=0.70\) and \(L_{\mathrm{con}}=0.30\), respectively, an adaptation period \(T_{\mathrm{adapt}}=100\) ms, and a radius step \(\Delta R=25\) m. For the cross-scheduler comparison, all six schedulers (\gls{edf}, \gls{pf}, \gls{rr}, \gls{fcfs}, Hybrid \gls{ml}-\gls{edf}, Weighted \gls{ml}-\gls{edf}) are evaluated across all ten scenarios. In all cause decomposition analyses, \gls{edf} scheduler is used as a reference. Additionally, a beam-sweep sensitivity analysis is conducted by varying $T_{\mathrm{sw}}$ within the range of $[10, 40]\,\mathrm{ms}$ for scenario S6. An antenna-count sweep is also performed, varying $M$ over the values $\{64, 128, \ldots, 4096\}$ with a fixed $B=256$ and $U=1000$.
\begin{table}
    {\small
    \centering
    \begin{tabular}{p{0.16cm}p{0.85cm}p{0.55cm}p{0.5cm}p{0.6cm}p{3.75cm}}
    \hline
    \textbf{ID} & \textbf{Antenna (M)} & \textbf{Beams (B)} & \textbf{Users (U)} &
    \textbf{Density} & \textbf{Mission Context} \\
    \hline\hline
    S1  & $64$   & $32$   & $100$  & Mod.  & Forward operating base \\
    S2  & $256$  & $64$   & $100$  & Mod.  & Tactical \gls{hq} node \\
    S3  & $256$  & $128$  & $100$  & Mod.  & Expanded \gls{hq} element \\
    S4  & $256$  & $64$   & $1000$ & High  & Small \gls{uav} swarm \gls{c2} \\
    S5  & $256$  & $128$  & $1000$ & High  & Expanded \gls{uav} swarm \gls{c2} \\
    S6  & $1024$ & $256$  & $1000$ & Extr. & Dense urban \gls{c2} hub \\
    S7  & $1024$ & $512$  & $1000$ & Extr. & Dense urban \gls{c2} (aug.) \\
    S8  & $2048$ & $1024$ & $1000$ & Extr.& Hyper-dense sensor net \\
    S9  & $4096$ & $2048$ & $1000$ & Extr. & Hyper-dense + high mobility \\
    S10 & $4096$ & $4096$ & $1000$ & Extr. & Full hyper-massive array \\
    \hline
    \end{tabular}
    \caption{Scenarios for the Decomposition Analysis.}
    \label{tab:scenarios}}
    \vspace{-1em}
\end{table}

\begin{figure}[!t]
    \centering
    \includegraphics[width=0.92\columnwidth]{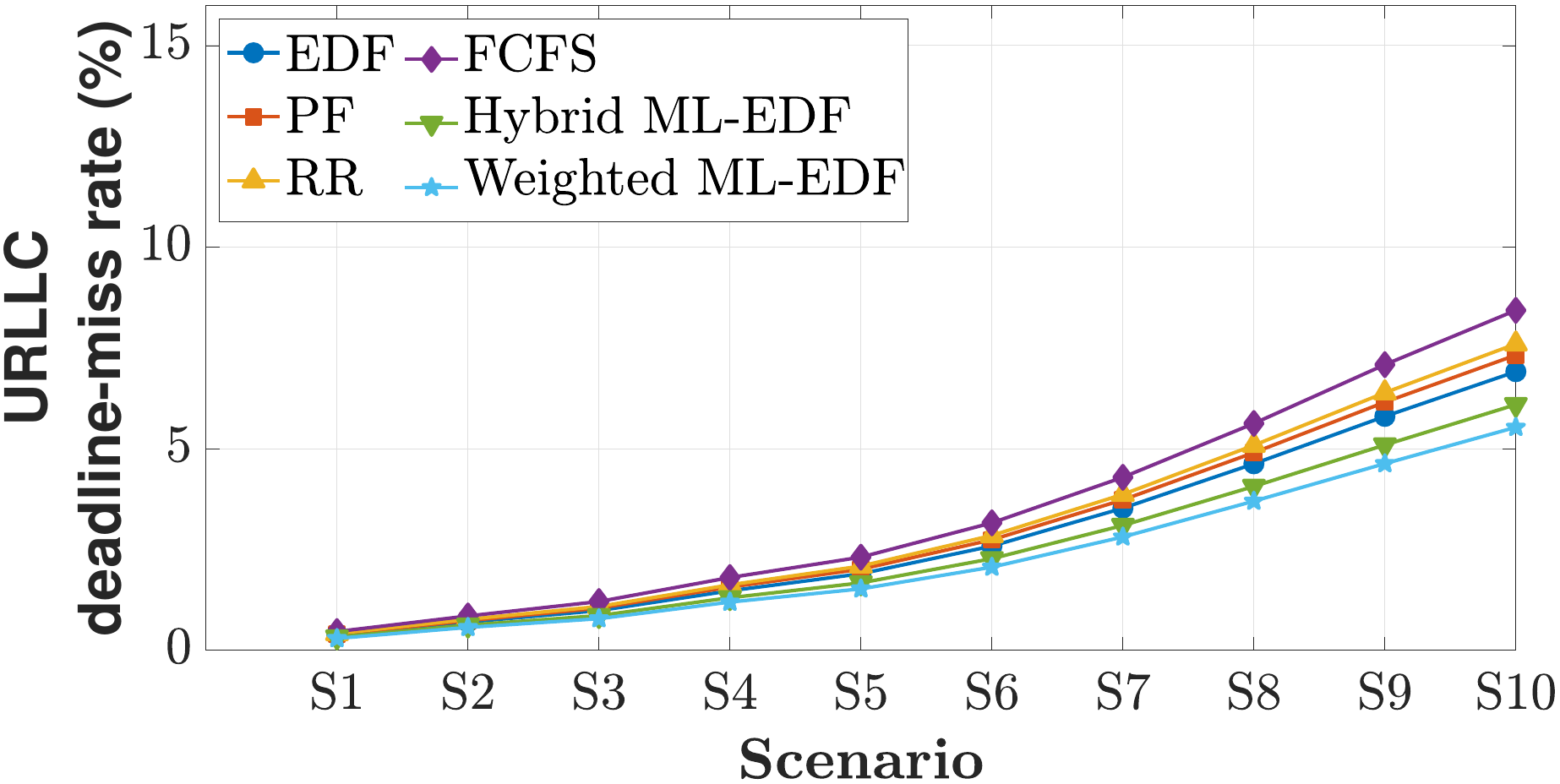}
    \caption{Aggregate \gls{urllc} deadline-miss rate by scenario.} 
    \label{fig:aggregate_miss_rate}
    \vspace{-1.25em}
\end{figure}

Fig.~\ref{fig:aggregate_miss_rate} shows the aggregate \gls{urllc} deadline-miss rate across all six schedulers. Under \gls{edf}, the miss rate increases gradually from $0.38\,\%$ in S1 (forward operating base) to $6.91\,\%$ in S10 (full hyper-massive array), representing an approximately $18$-fold increase. The ranking of the schedulers remains consistent across scenarios: Weighted \gls{ml}-\gls{edf} achieves the lowest miss rate at every operating condition, while \gls{fcfs} consistently has the highest. At S10, the best-to-worst scheduler gap reaches $2.90$ percentage points. Hybrid \gls{ml}-\gls{edf}, \gls{edf}, \gls{pf}, and \gls{rr} fall between these extremes, with their spread increasing from about $0.08$ percentage points in S1 to $1.51$ percentage points in S10. This widening is most pronounced at higher densities (S8-S10), indicating that scheduler choice becomes more critical as deployment density rises. However, miss rates remain substantial in the highest-density scenarios (S8-S10), even under the best-performing scheduler, indicating that scheduler selection alone is insufficient.

\begin{figure}[!t]
    \centering
    \includegraphics[width=0.92\columnwidth]{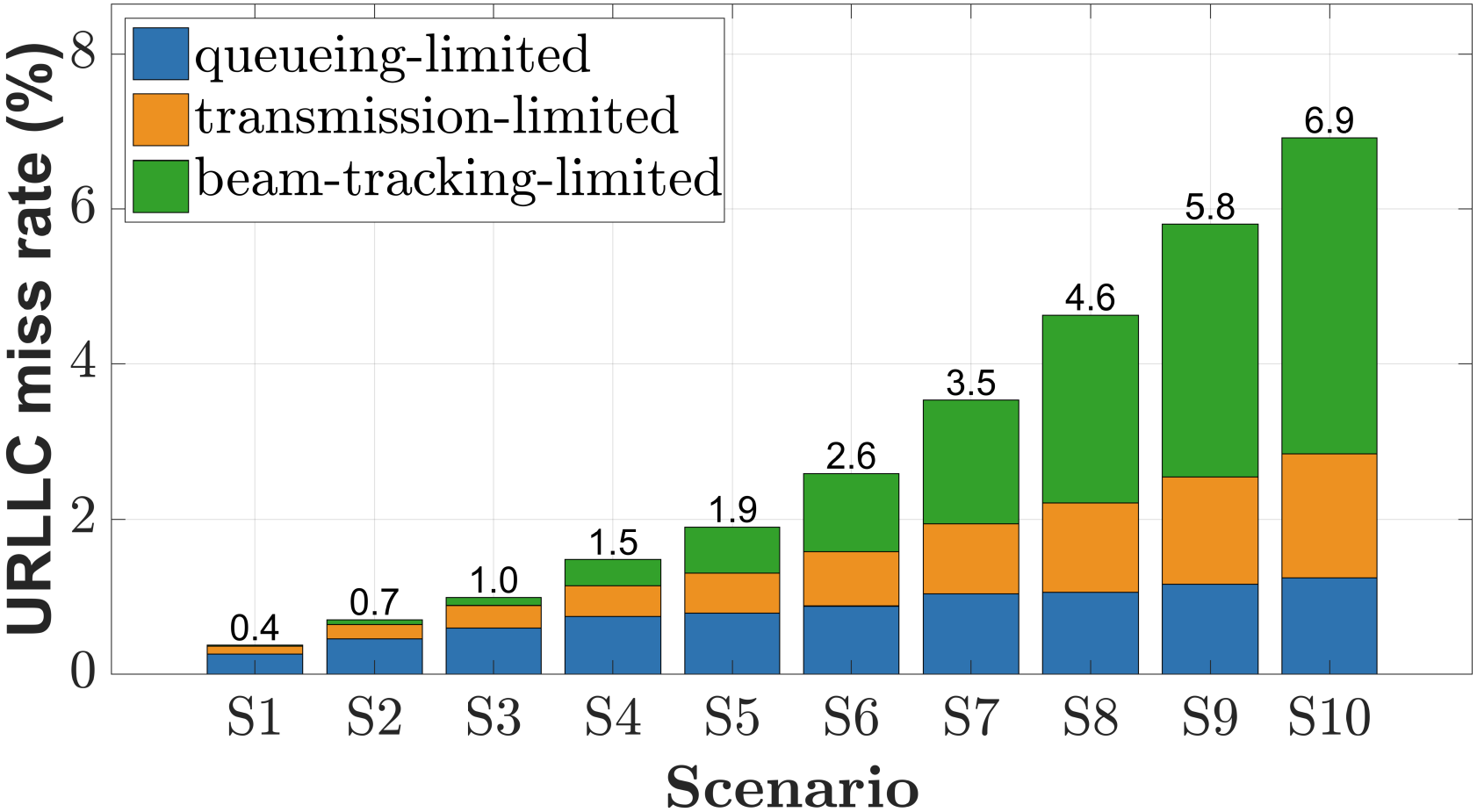}
    \caption{\gls{urllc} miss rate decomposed into Q (queueing), T (transmission), and B (beam-tracking) components per scenario under \gls{edf}. Total miss rate is annotated above each bar.}
    \label{fig:qtb_decomposition_edf}
    \vspace{-1.25em}
\end{figure}

Fig.~\ref{fig:qtb_decomposition_edf} illustrates the per-cause decomposition under \gls{edf} across all ten scenarios and reveals three significant operational findings. First, the dominant cause of misses changes with deployment density. 
In scenarios S1 to S3 (tactical nodes with $M\leq256$ and $U=100$), Q-misses account for approximately $61$-$69\,\%$ of total misses. This indicates that the system is constrained by beam resources, resulting in most packets failing before the scheduler can process them. Therefore, for operators at forward operating bases and \gls{hq} elements, adding beam or spectral resources, rather than improving beam tracking, is the primary strategy. 
At S6, where $M=1024$ in dense urban C2, the Q- and B-components become comparable at approximately $34\,\%$ and $39\,\%$ of misses, respectively, and by S7, B becomes clearly dominant over Q. 
In scenarios S8 to S10, with $M\geq2048$ and hyper-massive arrays, B clearly dominates, accounting for approximately $52$-$59\,\%$ of total misses. This is due to narrower beams ($\propto 1/\sqrt{M}$) becoming highly sensitive to the angular mobility of \gls{uav}s and ground vehicles. 
Second, the T-component consistently contributes approximately $23$-$29\,\%$ of total misses across the evaluated values of $M$, $B$, and deployment density. Thus, transmission-limited failures remain comparatively stable across the studied scenarios, forming a persistent baseline contribution. 
Third, the absolute contribution from the B-component increases significantly, growing by more than two orders of magnitude from $0.014\,\%$ in scenario S1 to $4.08\,\%$ in scenario S10. In contrast, the Q-component grows more modestly from $0.26\,\%$ to $1.24\,\%$. 
This confirms that the overall increase in miss rates observed across the scenarios is primarily driven by beam-tracking failures rather than scheduling congestion.

\begin{figure}[!t]
    \centering
    \includegraphics[width=0.92\columnwidth]{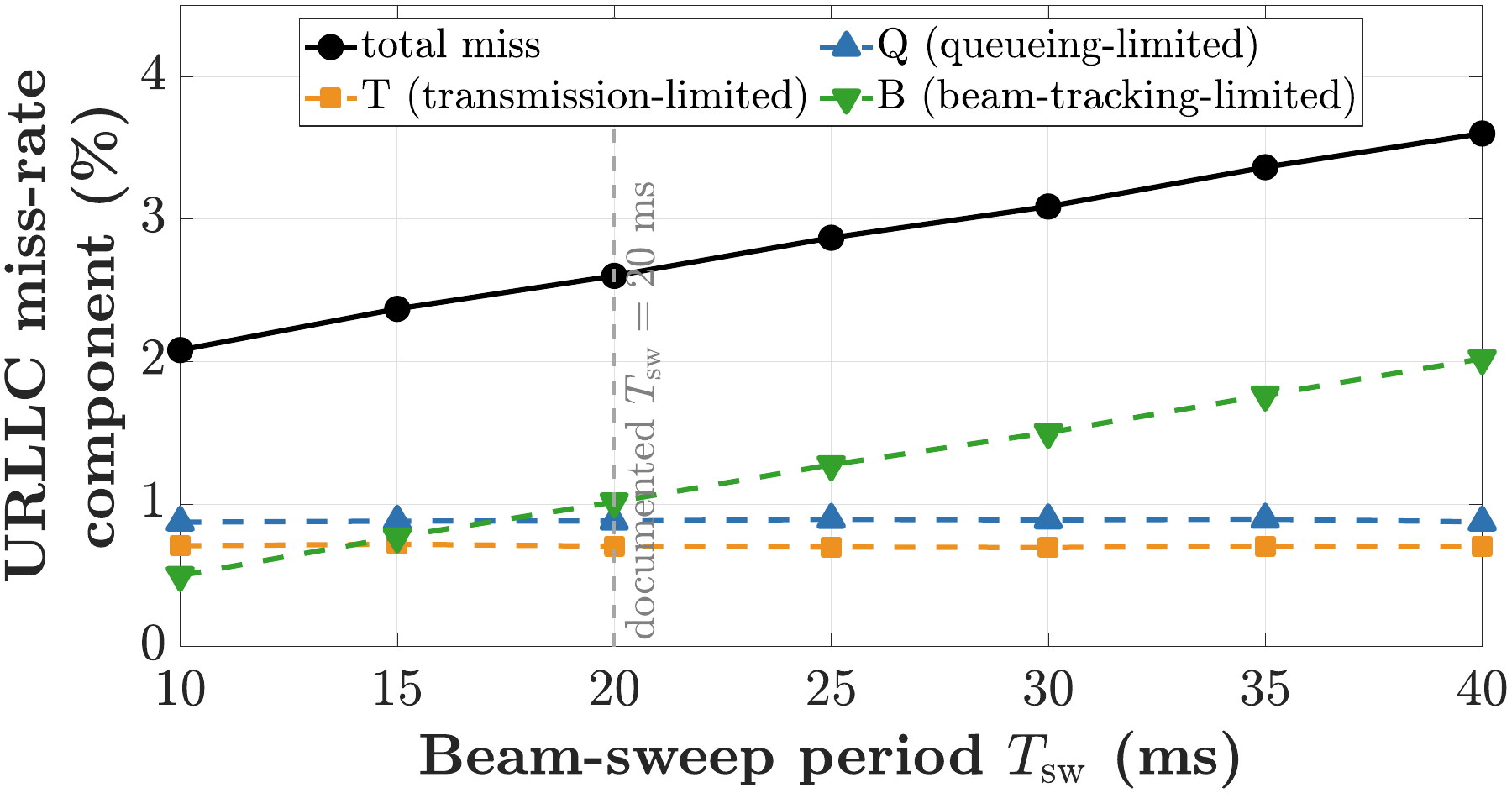}
    \caption{Per-cause miss components as a function of beam-sweep period $T_{\mathrm{sw}}$ at scenario S6.}
    \label{fig:tsw_sweep_qtb}
    \vspace{-1em}
\end{figure}

To isolate the beam-management lever, we swept $T_{\mathrm{sw}}$ from $10\,\mathrm{ms}$ to $40\,\mathrm{ms}$ in scenario S6, as shown in Fig.~\ref{fig:tsw_sweep_qtb}. 
The B-component scales approximately linearly with $T_{\mathrm{sw}}$, increasing from about $0.50\,\%$ at $10\,$ms to approximately $2.02\,\%$ at $40\,$ms, a roughly fourfold rise. In contrast, the Q component remains stable around $0.9\,\%$, and the T component stays at about $0.7\,\%$. This selectivity indicates that $T_{\mathrm{sw}}$ can be adjusted to reduce beam-tracking failures with minimal effect on the Q and T components. 
Notably, reducing $T_{\mathrm{sw}}$ from $20\,\mathrm{ms}$ to $10\,\mathrm{ms}$ cuts B-misses by over $50\,\%$. 
This suggests a mobility-adaptive sweep policy in which higher-mobility users are assigned shorter \(T_{\mathrm{sw}}\) to reduce beam-tracking failures, while lower-mobility users can retain longer sweep periods to limit unnecessary beam-sweeping overhead.

\begin{figure}[!t]
    \centering
    \includegraphics[width=0.92\columnwidth]{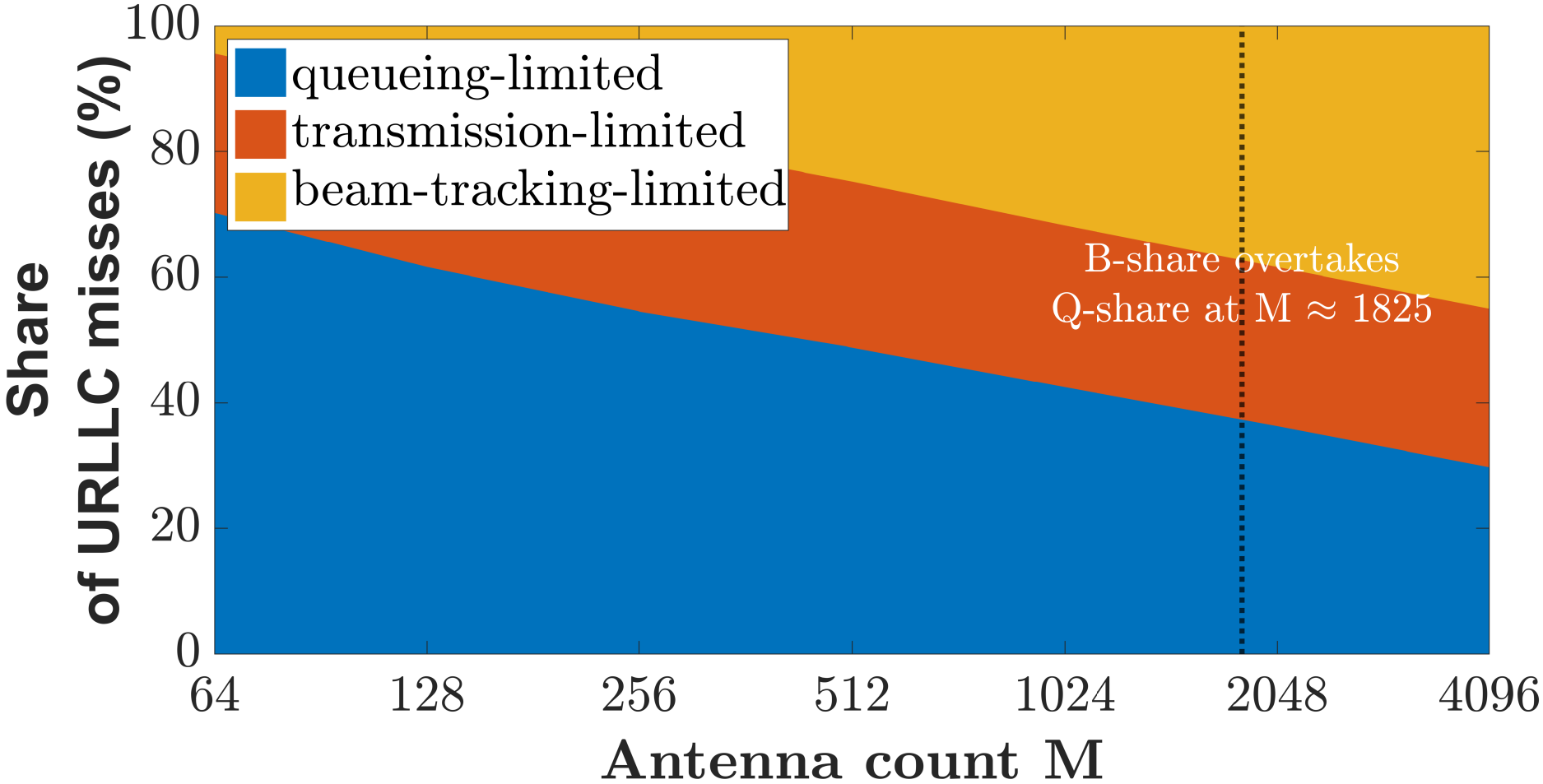}
    \caption{Stacked-area plot of cause shares as a function of antenna count $M$ ($B=256$ fixed).}
    \label{fig:antenna_cause_shares}
    \vspace{-1.25em}
\end{figure}

Fig.~\ref{fig:antenna_cause_shares} provides a comprehensive view of the structural dynamics by varying $M$ while keeping $B$ fixed at $256$ and $U$ at $1000$ across the entire antenna range. The Q-share declines consistently, dropping from about $70\,\%$ at $M=64$ to about $30\,\%$ at $M=4096$. In contrast, the B-share rises from approximately $4\,\%$ to $45\,\%$, with interpolation between simulated antenna-count points placing the Q-B crossover at $M\approx1825$ under the studied conditions. The T-share remains comparatively stable throughout, ranging from approximately $25\,\%$ to $27\,\%$, consistent with the relatively stable T-share observed in Fig.~\ref{fig:qtb_decomposition_edf}. Although increasing $M$ enhances array gain, the T-share remains comparatively stable in the controlled sweep. However, the narrower beamwidth associated with larger arrays makes the system increasingly susceptible to angular drift and beam misalignment.

\begin{figure}[!t]
    \centering
    \includegraphics[width=0.92\columnwidth]{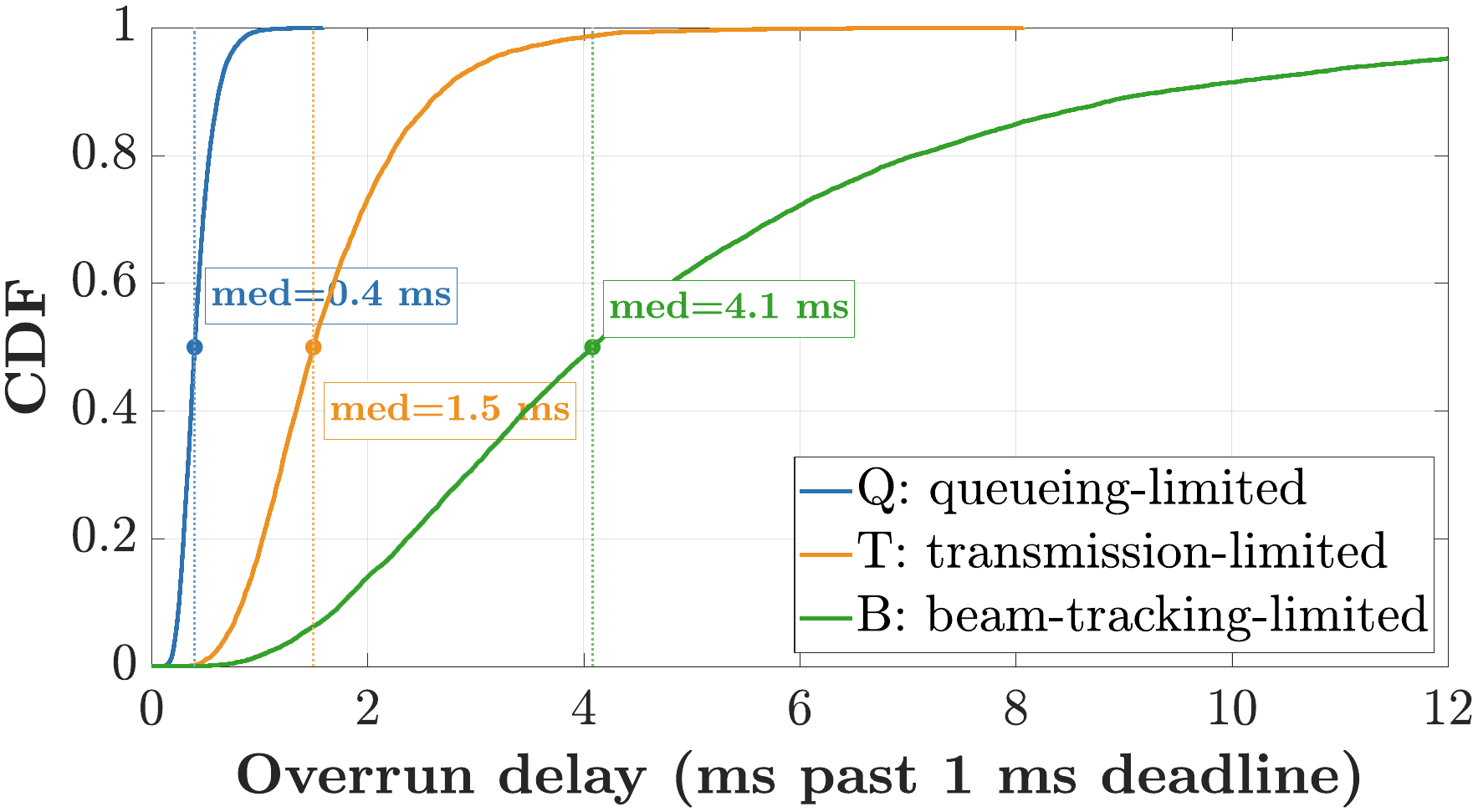}
    \caption{Empirical \gls{cdf} of overrun delay (ms past the $1$\,ms \gls{urllc} deadline) for missed packets, by cause class.}
    \label{fig:Overrun_CDF_By_Cause}
    \vspace{-1.25em}
\end{figure}

The point at which the B-share overtakes the Q-share represents a critical design transition. Below this threshold, queueing failures are the primary concern, suggesting that improvements in scheduling or capacity will most effectively reduce miss rates. Beyond this point, beam-tracking failures become more significant, indicating that optimizing the scheduler alone may yield diminishing returns, and improvements in beam-layer elements, such as tracking accuracy, codebook refinement, and angular prediction, become increasingly important. For military planners tasked with specifying antenna apertures for \gls{c2} nodes or coordinating multiple \gls{uav}s, this threshold is crucial. It clearly delineates where enhancements in beam-tracking capabilities should be integrated with increases in array size \cite{larsson2014,bjornson2025gigantic}.
Finally, Fig.~\ref{fig:Overrun_CDF_By_Cause} illustrates the severity of each cause class by presenting empirical \glspl{cdf} of overrun delay. This delay refers to the time that a missed packet exceeds the $1\,\mathrm{ms}$ \gls{urllc} deadline. The three distributions are distinctly separated in terms of both location and tail behavior. Q-misses have a median overrun of $0.4\,$ms, with the \gls{cdf} reaching approximately $0.98$ by $0.8\,$ms and continuing toward $1.0$ thereafter.
Thus, most Q-miss overruns could be accommodated by a sub-millisecond latency-budget extension where such relaxation is permissible. 
T-misses have a higher median overrun of $1.5\,$ms, with the \gls{cdf} reaching approximately $0.97$ around $3.5\,$ms. In contrast, B-misses are qualitatively different, exhibiting a median overrun of $4.1\,\mathrm{ms}$. 
The B-miss \gls{cdf} remains unsaturated at $12\,$ms, making such overruns difficult to accommodate within practical application-layer latency budgets. This suggests that beam-tracking failures can become a limiting mechanism for safety-critical \gls{c2} links and are more appropriately addressed at the physical and \gls{mac} layers. 
\section{Mission-Critical Implications}
\label{sec:implications}
The decomposition process transforms a complex aggregate miss rate into a three-way diagnostic that clearly identifies the system layer requiring attention with minimal additional instrumentation. 
Table~\ref{tab:mapping} summarizes how each failure cause maps to the observed operational regime and the recommended remediation.
\begin{table*}
    \begin{tabular}{p{0.75cm}p{4cm}p{12cm}}
    \hline
    Cause &  Regime & Primary Remedy\\
    \hline\hline
    Q & $M$ $\leq 256$, U$\leq100$ (forward bases, expeditionary \gls{hq}) &  Implement urgency-aware scheduling (such as \gls{edf} and its variants) and reserve dedicated \gls{urllc} beam slots. \\
    T & $23$-$29\,\%$ across evaluated \gls{edf} scenarios & Enhance \gls{sinr} through additional antennas, wider bandwidth, and smaller cell radius; utilize frequency agility and interference cancellation in jamming conditions.\\
    B & $M \gtrsim 1825$ under studied conditions
    & Reduce $T_{\mathrm{sw}}$ (halving it to $10\,\mathrm{ms}$ can decrease B-misses by over $50\,\%$); expand the beam codebook for finer angular resolution; and deploy \gls{ekf}-based angular predictive tracking.\\
    \hline
    \end{tabular}
    \caption{Mapping of Failure Causes to Operational Regimes and Recommended Remediation Strategies.}
    \label{tab:mapping}
    \vspace{-1.2em}
\end{table*}
Three key points emerge from this mapping: First, in the Q-dominated regime, investing in beam tracking yields minimal benefits, as B-misses constitute a small portion of total failures. Hence, the correct levers to focus on are scheduling and spectrum management. Second, the T-component remains comparatively stable under the evaluated benign conditions but could increase under adversarial jamming. In such cases, link protection measures such as spread spectrum, rapid power control, and receiver-side cancellation become appropriate remediation mechanisms. Third, B-misses remain unsaturated at $12\,$ms, making physical- and \gls{mac}-layer remediation more practical than application-layer latency relaxation. Operators should therefore select remediation according to the dominant failure cause rather than defaulting to scheduler tuning. 
\section{Limitations}
\label{sec:limitations}
The classifier assigns each miss to a singular proximate cause. However, in practice, packets frequently experience multiple contributing factors. For instance, a \gls{urllc} packet delayed by queueing congestion might 
be placed into a beam-misaligned slot, as either cause alone would be sufficient to create a miss. 
While the proximate-cause attribution is algorithmically clear, assigning each miss to a single category necessarily loses information about concurrent contributing mechanisms. A combinatorial decomposition encompassing all cause subsets (Q-only, T-only, B-only, Q${\cap}$B, T${\cap}$B, etc.) remains for future research and would require expanding the classifier to monitor multiple concurrent active conditions per packet. The misalignment predicate relies on ground-truth angular position, which is accessible in simulations but not on deployed systems. An alternative proxy based on received-signal-quality was introduced in Section~\ref{sec:framework}. However, its classification accuracy under conditions of multipath fading, co-channel interference, and adversarial jamming necessitates further experimental validation on a real \gls{mmwave} testbed before the runtime classifier can be deemed production-ready.

The empirical findings depend on the \gls{3gpp} TR~$38.901$ urban-microcell channel model and pre-defined simulation parameters, including $100$ Monte Carlo trials for each scenario. Military deployments present key differences: contested spectrum, variable interference, electronic warfare, and operationally driven platform trajectories may significantly alter cause-share distributions compared to those reported here. 
Notably, the crossover at \(M\approx1825\) and the T-component baseline of $23$-$29\,\%$ are based on the \gls{3gpp} channel model and the modeled mobility conditions. Adversarial \gls{rf} environments could markedly increase the T-component, shifting the crossover point and modifying the regime boundaries outlined in Section~\ref{sec:results}. Evaluation under adversarial channel models and military-relevant mobility patterns is essential for future work.
\section{Conclusion}
\label{sec:conclusion}
We introduced a decomposition framework that partitions the \gls{urllc} deadline-miss rate into three categories: queueing-limited, transmission-limited, and beam-tracking-limited. This includes formal definitions, a low-overhead per-packet classifier, and per-cause overrun-delay characterization. Analyzed across ten scenarios, from small tactical nodes to hyper-massive \gls{uav} coordination platforms, this framework yields several key findings with direct operational impact. 
Under the studied conditions, interpolation places the Q-B crossover near $M\approx1825$, beyond which beam-tracking improvements become increasingly important relative to scheduler-focused remediation. The beam-sweep period $T_{\mathrm{sw}}$ serves as a direct lever for the B-component: halving it from $20\,\mathrm{ms}$ to $10\,\mathrm{ms}$ reduces B-misses by over $50\,\%$ with minimal impact on Q or T. 
The T-component remains comparatively stable across the evaluated \gls{edf} scenarios but could increase under adversarial interference. 
The B-miss \gls{cdf} remains unsaturated at $12\,$ms, whereas the Q-miss \gls{cdf} approaches $1.0$ near $0.8\,$ms, making modest latency-budget relaxation substantially more practical for Q-misses. These insights transform a complex aggregate miss-rate metric into a three-part diagnostic tool, enabling military network operators to efficiently attribute failures in high-density, high-mobility environments with minimal additional instrumentation. 
\begin{spacing}{0.92}
\bibliographystyle{IEEEtran}
\bibliography{references}
\end{spacing}
\end{document}